\documentclass[usenatbib]{mn2e}
\usepackage[totalwidth=170mm, totalheight=240mm, paperwidth=8.5in,  paperheight=11in]{geometry}

\usepackage{epsfig}
\usepackage{amsmath}

\newcommand{\degree}{\ensuremath{^\circ}}

\def\gtap{\raisebox{-.55ex}{\rlap{$\sim$}} \raisebox{.4ex}{$>$}}
\def\gsim{\mathrel{\gtap}}

\title{Flux calculations in an inhomogeneous Universe: weighting
a flux-limited galaxy sample}

\author[Hylke Koers and Peter Tinyakov]
{Hylke B. J. Koers$^{1}$\thanks{E-mail: hkoers@ulb.ac.be} and
Peter Tinyakov$^{1,2}$\thanks{E-mail: Petr.Tiniakov@ulb.ac.be} \\
$^{1}$Service de Physique Th\'eorique, Universit\'e Libre de Bruxelles (U.L.B.), CP225, Bld. du Triomphe, B-1050 Bruxelles, Belgium\\
$^{2}$Institute for Nuclear Research, 60th October Anniversary Prospect 7a, 117312, Moscow, Russia}

\begin{document}
\pagerange{\pageref{firstpage}--\pageref{lastpage}} \pubyear{2009}
\maketitle
\label{firstpage}
\begin{abstract}
Many astrophysical problems arising within the context of
ultra-high energy cosmic rays,
very-high energy  gamma rays or neutrinos,
require calculation of the flux produced
by sources tracing the distribution of galaxies in the Universe. We
discuss a simple weighting scheme, an application of the method introduced
by Lynden-Bell in 1971,
that allows the calculation of the flux
sky map directly from a flux-limited galaxy catalog without cutting a
volume-limited subsample. Using this scheme, the galaxy distribution can be modeled
up to large scales while representing the distribution in the nearby Universe
with maximum accuracy.
We consider fluctuations in the flux map arising
from the finiteness of the galaxy sample. We show how these fluctuations
are reduced by the weighting scheme and discuss how the remaining fluctuations
limit the applicability of the method.
\end{abstract}

\begin{keywords}
methods: miscellaneous, catalogues, large-scale structure of Universe,
galaxies: luminosity function, cosmic rays
\end{keywords}

\section{Introduction}
\label{sec:intro}

Recent developments in multi-wave\-length/multi-mes\-sen\-ger
observational techniques often make it desirable to calculate the
angular distribution of a diffuse flux expected from sources with a
given spatial distribution. The predicted flux distribution may then be
used for source identification, estimation of the background,
etc. Necessity for such a calculation arises in the context of
ultra-high energy cosmic rays (UHECRs), neutrino physics, as well as
gamma-ray astronomy.

If the sources are extragalactic, their space distribution can be
derived from the matter distribution in the Universe. The latter can be
inferred from galaxy surveys, e.g.
\citet{2008ApJS..175..297A, 2006AJ....131.1163S, 2005PASA...22..277J}.
Good distance determination is required
to reconstruct the spatial mass distribution. Special techniques
have been developed to minimize the impact of distance errors and to
suppress the short-scale noise (see, e.g., \citet{Erdogdu:2006nd}
and references therein).

The problem of flux calculation has a number of features that make it
different from (and easier than) reconstruction of the full
three-dimensional mass distribution: \emph{(i)} only a two-dimensional
projection of the three-dimensional distribution is needed;
\emph{(ii)} contributions of remote sources are suppressed by the
geometrical factor $r^{-2}$ and, in many cases, by the flux attenuation due to
interactions with the ambient matter; \emph{(iii)} the smaller
amplitude of inhomogeneities at larger scales makes the contribution
of remote sources essentially isotropic; only the overall normalization
of such an isotropic part has to be calculated. These simplifications
result in weaker requirements on the quantity and quality of
astronomical data in flux calculations, which makes it advantageous to
by-pass the reconstruction of matter density and calculate the flux
distribution directly from the galaxy catalogs. Accurate results may
be achieved with substantially smaller input.

Both in the context of mass distribution and in flux calculations, a
crucial requirement is completeness of the underlying galaxy
catalog. That is, a volume-limited sample is needed which includes
all galaxies of a certain kind within a given volume. On the contrary, a
natural product of an astronomical survey is a flux-limited sample
that contains all galaxies up to certain {\em apparent} magnitude
as set by the instrumental sensitivity and observation
time. Volume-limited samples may be obtained from a flux-limited
sample by cutting away objects that are further than a given distance and
dimmer than a certain {\em absolute} magnitude, chosen in such a manner that the
resulting sample is complete. 

To model adequately the source distribution in the Universe, one
requires a galaxy catalog that \emph{(i)} extends to
sufficiently large distance (large enough that  the
Universe can be approximated as homogeneous beyond that), and \emph{(ii)} accurately
represents the distribution of matter on small distance scales. For
any single volume-limited subsample, these requirements work in
opposite directions: the first calls for a large volume, while the
second calls for a small volume. To resolve this conflict
one could consider combining two volume-limited samples,
a sparse one to cover large distances
and a dense one to cover the nearby region in more detail.
In this paper we discuss a technique -- termed ``sliding-box'' technique --
which generalizes  this idea by
combining many  volume-limited samples.
Up to a specified limiting distance the whole flux-limited
catalog is used in this process, so that
close-by structures are mapped out with maximum detail
by the dim objects in the original catalog.
The crux of the construction is 
in an appropriate weighting scheme:
We weigh galaxies in the flux-limited sample
in a distance-dependent way so that the progressive incompleteness at
large distances is compensated by the increasing weight of each
galaxy. These weights, essentially representing
the luminosity per galaxy,
can be naturally incorporated in
flux computation algorithms
which inherently use some sort of
weights to account for the fact that remote sources
produce less flux than nearby ones.

The sliding-box technique is essentially a method to deal with
the fact that galaxy catalogs do not contain an infinite number of galaxies.
The finiteness of a galaxy catalog unavoidably leads to
fluctuations in flux predictions.
The sliding-box technique strongly reduces these fluctuations
by efficient use of the available data.
Nevertheless, the remaining fluctuations may still be
large enough to spoil the accuracy of flux maps modeled
from a galaxy catalog.
We will address this problem in detail and present a
criterion for the applicability of the sliding-box method.

An an illustration we will apply 
the sliding-box technique
to a subset of the 2 Micron All-Sky Redshift
Survey (2MRS) \citep{2MRS},
a flux-limited sample of galaxies with observed $K_s$-magnitude
$m \leq 11.25$ that contains measured redshifts
for all but a few galaxies.\footnote{\label{footnote1} This subset was kindly provided to
us by John Huchra. Tailored to model the distribution of galaxies in
the field of view of a northern-hemisphere cosmic-ray experiment, it
does not cover the galactic plane with $|b|<10\degree$, $b$ being the
galactic latitude, nor the region with $\delta < -30\degree$, $\delta$ being
the declination in J2000 equatorial coordinates.}
We would like to stress, however, that the technique is completely general
and can be applied to any flux-limited galaxy sample.

It is worth noting that the issue addressed in this paper is related to the well-known astronomical
problem of reconstructing the luminosity function from a quasar or galaxy sample that is
limited in apparent magnitude (see, e.g., \citet{1977AJ.....82..861F, 1997AJ....114..898W}
for a comparison between different methods and references). 
The sliding-box scheme discussed in this work is an 
application
of the $C^{-}$-method that was proposed by 
\citet{1971MNRAS.155...95L}, and further developed by \citet{1974MNRAS.166..281J,
1987MNRAS.226..273C,1988MNRAS.232..431E}, to reconstruct the quasar luminosity function.
Our formulation of the scheme is tailored
for flux calculations, in keeping with the aim of the present study.

The rest of this paper is organized as follows. In section
\ref{sec:slid-box-weight} we discuss the sliding-box
technique and present an efficient implementation scheme. We also discuss
the connection
between the weights associated with the algorithm on one hand
and the luminosity and selection functions on the other.
Section \ref{sec:fluctuations} is concerned with the effect of
fluctuations on
model fluxes due to the finite size of a galaxy sample.
In section~\ref{section:2MRS}, as an example, we apply the sliding-box
method to model the flux
of UHECR protons with energies above 60~EeV from the 2MRS sample.
We summarize our work in section \ref{sec:summary}.

\section{The sliding-box weighting scheme}
\label{sec:slid-box-weight}
\subsection{Combining two volume-limited samples}

\begin{figure}
\includegraphics[angle=0, width=8cm]{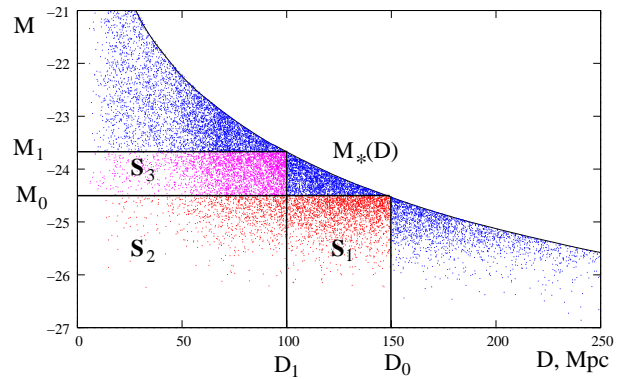}
\caption{\label{fig:cutting-samples} Two nested volume-limited
  subsamples of a flux-limited sample.}
\end{figure}
To illustrate the general idea of the sliding-box technique, consider a
flux-limited sample that is complete to a given apparent magnitude
$m_0$. On the $M-D$ plane, where $M$ is the absolute magnitude and $D$
is the distance, this sample occupies the populated  region in
Fig.~\ref{fig:cutting-samples}. The apparent magnitude $m$ of a source
is a function of its absolute magnitude and distance, $m=m(M,D)$. For a
given absolute magnitude, $m$ increases with distance and reaches the
limiting value $m_0$ at a distance $D$ satisfying
$m(M,D)=m_0$. This determines the line $M_*(D)$, the boundary of the
populated region in Fig.~\ref{fig:cutting-samples}. Beyond this line
the objects are too dim and the completeness of the sample cannot be
guaranteed.

At a given distance $D_0$, only objects with absolute magnitude $M<M_0
= M_*(D_0)$ are sufficiently bright to be included in the flux-limited
catalog. Galaxies that are closer than $D_0$ and brighter than $M_0$
form a volume-limited sample. These are objects in regions marked
with $S_1$ and $S_2$ in Fig.~\ref{fig:cutting-samples}. The
completeness of this subsample follows from the completeness of the
original flux-limited catalog.

It is clear from figure~\ref{fig:cutting-samples} that at small
distances the volume-limited sample $S_1+S_2$ contains only a fraction
of available galaxies which may be insufficient to represent
accurately the details of the matter distribution. To improve the
situation, one may construct a denser volume-limited sample
corresponding to a smaller distance $D_1$ (the sample $S_2+S_3$ on
Fig.~\ref{fig:cutting-samples}). When calculating the flux one may use
galaxies from $S_2+S_3$ at distances $D<D_1$ and galaxies from $S_1$
at $D_1<D<D_0$. The luminosity of a given volume is determined by the
number of galaxies in the sample $S_1+S_2$ contained in that
volume. At distances $D<D_1$, the same luminosity may be represented
in a greater detail by galaxies from $S_2+S_3$ provided they are
assigned smaller ``weight'', that is, luminosity per galaxy. If the
galaxies in the sparse sample have a weight $w_0$ each, the galaxies
in the dense sample $S_2+S_3$ should be weighted with
\begin{equation}
w_1 = {S_2\over S_2+S_3}w_0.
\label{eq:w-first-def}
\end{equation}
Here and below we use the same letter to denote the sample and the
number of galaxies in the sample. At distances $D<D_1$, the total
weight in the sparse and dense samples is the same,  $S_2 w_0$. The
difference is that in the dense sample it is distributed among a larger
number of galaxies, and hence the spatial distribution of matter is
represented more accurately.

Several volume-limited samples may be combined in the same way. In the
limit of an infinite number of nested volume-limited samples one
arrives at the ``sliding-box'' weighting scheme described now.
This scheme is essentially an implementation of the $C^-$-method,
proposed by  \citet{1971MNRAS.155...95L}, applied to distance $D$ and
magnitude $M$.

\subsection{Sliding-box scheme}
Imagine a variable
rectangular ``sliding box'' with one corner fixed at 
zero distance and minimum $M$ 
(the lower-left corner in figure~\ref{fig:cutting-samples}) and the
opposite corner moving along the line $M_*(D)$. At any given position
the box defines a volume-limited sample. One starts at some maximum
distance $D_{\rm max}$; galaxies that are further than $D_{\rm max}$
are disregarded, i.e. assigned a zero weight (this is the part of
the catalog that is lost). The current weight is set to, say, 1. Now
the free corner of the sliding box is moved towards smaller
distances. Each time a galaxy exits the box through its vertical edge
it is assigned the current weight. Each time a galaxy enters the box
through the horizontal edge, the current weight is multiplied by
$N/(N+1)$, where $N$ is the current number of galaxies in the
box. When the procedure is finished, all the galaxies at $D<D_{\rm
max}$ have been assigned a weight.

\begin{figure}
\includegraphics[angle=270, width=8cm]{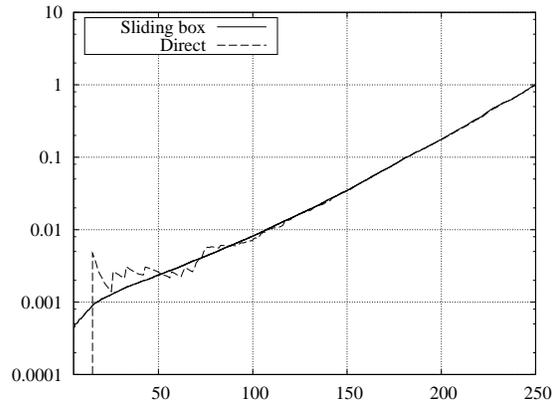}
\caption{\label{fig:weights} Galaxy weights as a function of distance,
obtained with the sliding-box algorithm and with a direct computation.}
\end{figure}

The main asset of the sliding-box scheme is that the
weight at a given scale $D$ is computed from a
volume-limited sample corresponding to distances just slightly larger
than $D$; this sample has the maximum available number of galaxies and
hence the smallest fluctuations. 
To demonstrate the accuracy of the  sliding-box
method, consider a direct computational scheme in which
the weight at distance $D$ is determined from a volume-limited sample
up to $D$ and a volume-limited sample up to $D_{\rm max}$
(i.e., without refining the weights at intermediate
distance as is done in the sliding-box method).
The direct scheme and the sliding-box method are equivalent
in the limit of infinite galaxies in the original sample.
Given a finite number of galaxies, however,
large statistical fluctuations will show up in the direct
computational scheme  due to  the sparsity of the 
volume-limited sample extending to $D_{\rm max}$ at small distances.
This can be seen clearly
in figure~\ref{fig:weights}, which shows the weights as a
function of distance for all galaxies in the 2MRS sample computed by
the sliding-box technique and by  direct computation.

\subsection{Relation to luminosity and selection functions}
\label{sec:weight-function-vs}

The weight defined by the sliding-box method is related to 
the galaxy luminosity distribution and to the selection
function characterizing the flux-limited galaxy sample.
The original construction by \citet{1971MNRAS.155...95L} was, in fact, aimed
at recovering the quasar luminosity function (see also \citealt{1974MNRAS.166..281J,
1987MNRAS.226..273C,1988MNRAS.232..431E}).
To clarify these relations
we consider the problem in general terms. We assume in
this section that all the samples are  very large so that a statistical
description applies.  For simplicity, we also assume
that the distribution of galaxies in luminosity is
space-independent\footnote{The flux non-uniformity arises at $z\ll 1$
where one can neglect the evolution of sources. The contribution from
regions $z\gsim 1$ may be non-negligible or even dominant; however,
this contribution is isotropic.}, that is the full distribution
factorizes into a spatial and a luminosity part. The number of
galaxies with magnitudes between $M$ and $M+dM$ at distances between 
$D$ and $D+dD$ is then expressed as a product of two factors, 
\[
dN = \lambda(M)dM \cdot\nu(D) D^2 dD.
\]
Let $N(D,M)$ denote the total number of galaxies within distance $D$
and brighter than $M$. It factorizes into a product of two cumulative
distributions:
\begin{subequations}
\label{eq:defNDM}
\begin{equation}
N(D,M) = N(D)L(M) \, ,
\end{equation}
where
\begin{equation}
N(D)  \equiv \int_{0}^D dD {D}^2 \nu(D) \, ;
\end{equation}
and
\begin{equation}
L(M)   \equiv  \int_{-\infty}^M d M  \lambda(M) \, .
\end{equation}
\end{subequations}
According to eq.~(\ref{eq:w-first-def}), the weight function $w(D)$ at
some distance $D<D_{\rm max}$ may be expressed as follows:
\begin{equation}
\label{eq:def:w}
w(D) = 
{N(D, M_*(D_{\rm max})) \over N (D, M_*(D))}=
{L(M_*(D_{\rm max})) \over L(M_*(D))} ,
\end{equation}
where we have normalized the  weights so that $w(D_{\rm max})=1$.
Equations \eqref{eq:defNDM} and \eqref{eq:def:w} imply that 
\begin{equation}
\label{eq:wLfunc}
w(D)^{-1}  \propto  \int_{-\infty}^{M_* (D)} \lambda(M') dM' \, .
\end{equation}
This relation can be inverted to read
\begin{equation}
\lambda (M) \propto 
\frac{w' (D_*(M))}{w (D_*(M))^2} 
\frac{d D_*(M)}{d M}  \, ,
\end{equation}
where $w'(D) = dw /dD$ and $D_*(M)$ is the inverse of $M_*(D)$ defined
by $M_*(D_*(M))=M$.  Hence we can directly infer the galaxy luminosity
distribution $\lambda(M)$ from the weight function $w(D)$ and {\em vice
versa}.

\begin{figure}
\includegraphics[angle=270, width=8cm]{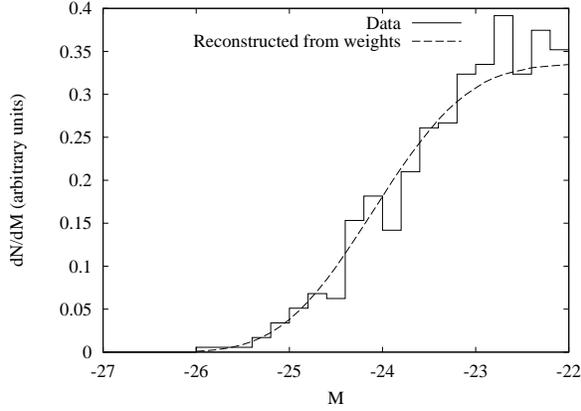}
\caption{\label{fig:distMplot} Luminosity distribution $\lambda(M)$ for a
volume-limited subsample of the 2MRS sample up to 30 Mpc.}
\end{figure}
Figure \ref{fig:distMplot} shows the luminosity distribution 
for a volume-limited subsample of the 2MRS sample
up to 30 Mpc derived
from the weights,
together with the distribution directly reconstructed from the data.  The
agreement is excellent.

\begin{figure}
\includegraphics[angle=270, width=8cm]{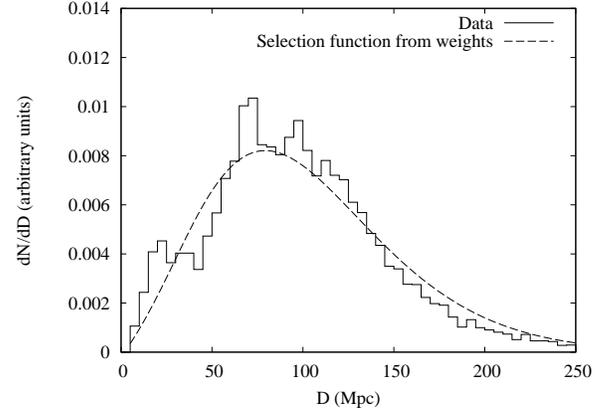}
\caption{\label{fig:selfunc}
Distribution $dN /dD$ of distances in the 2MRS sample compared
to expectations for a homogeneous universe. The model 
curve is proportional to $D^2 \phi_{\rm sel} (D)$, where
$\phi_{\rm sel}$ denotes the selection function.
}
\end{figure}

Now consider the relation between the weights and the selection
function $\phi_{\rm sel}(D)$ in a flux-limited sample. The total
number of galaxies within distance $D$ in an incomplete sample is
expressed in terms of the selection function as follows, 
\begin{equation}
n(D) = \int_0^D dD {D}^2 \nu(D) \phi_{\rm sel}(D).
\label{eq:N-of-D}
\end{equation}
Making use of the general
expressions in eqs.~(\ref{eq:defNDM}) one finds that
\begin{equation}
\label{eq:de:wi}
\phi_{\rm sel}(D) = w(D)^{-1} \cdot L(M_*(D_{\rm max})) \, ,
\end{equation}
where the last factor is just a normalization constant. The selection
function derived from the weights as given in eq.~\eqref{eq:de:wi} is
shown in figure \ref{fig:selfunc}. The model curve
is in fair agreement with the data, given that it is
derived under the assumption of a homogeneous Universe.

\section{Accuracy of flux predictions}
\label{sec:fluctuations}

In this section we address fluctuations 
associated with the finite number of galaxies
in flux maps modeled
from a galaxy catalog.
Although  the sliding-box
technique
efficiently suppresses these fluctuations, they still pose a potential
limitation to the applicability of the method. This makes it important
to quantify them, as we do in the following.

First we briefly discuss the construction of flux maps from a galaxy
catalog. We express the model flux from galaxy $i$ as follows:
\begin{equation}
\label{eq:defFi}
F_i = {F_0  w_i J(D_i) \over 4 \pi D_i^2} \, , 
\end{equation}
where $F_0$ is a normalization constant, $w_i$ is the weight assigned
to galaxy $i$ by the sliding-box technique, $D_i$ is the galaxy distance, and
$J(D)$ represents
the fraction of the integral flux from a source at distance $D$ that 
survives attenuation by redshift and interaction with the ambient matter.
The function $J(D)$ is different for UHECRs, neutrinos, and gamma rays;
$J(D)=1$  corresponds to no attenuation.
To keep the discussion general we do not specify $J(D)$ at this point.

Generally speaking, the model flux for a given direction on the sky is constructed
by adding and averaging the fluxes of individual sources 
close to the line of sight. This can be done in various ways, in particular
by dividing the sky into
bins or by employing a smearing routine that distributes single-source fluxes
over (part of) the sky. For the discussion of fluctuations the precise method
is not very important; 
the critical parameter is the solid angle $\Delta\Omega$
over which flux contributions of individual sources are averaged.

How large may fluctuations be in order not to spoil the accuracy of the
flux map? The answer to this question depends clearly on 
the purpose of the map and can thus not be answered in general.
A reasonable requirement,
which we will pursue in the following,
is that no significant
contribution to the flux within the solid angle $\Delta\Omega$ should come
from a single source, i.e.:\footnote{Alternatively, one could demand
that $F_i$ be smaller than the \emph{average} flux per solid angle $\Delta\Omega$. This criterion
would lead to equation~\eqref{eq:flucts-criterion-2} directly.}
\begin{equation}
{F_i\over \Delta F} \ll 1 \, ,
\label{eq:flucts-criterion}
\end{equation}
where $\Delta F$ denotes the total flux of
sources within $\Delta\Omega$.

With a few  simplifying assumptions, the condition \eqref{eq:flucts-criterion}
can be expressed in general terms.
We first approximate
$\Delta F \simeq F \Delta \Omega / \Omega$, where 
$F=\sum_i F_i$ stands for the total flux and
$\Omega$ is the total solid angle occupied by
the sample ($\Omega = 4 \pi$ in the case of complete sky coverage).
Equation~\eqref{eq:flucts-criterion} then reduces to:
\begin{equation}
{\Omega\over \Delta\Omega}{F_i\over F} \ll 1 \, .
\label{eq:flucts-criterion-2}
\end{equation}
To satisfy this requirement it is necessary to
have many sources in the solid angle $\Delta\Omega$. However, this may
be not sufficient because not all sources contribute the same
flux at Earth. The situation is thus complicated by the dependence
on distance $D$.
Introducing the
fraction $f(D)$ of the total flux produced by the sources closer than
$D$, and the number of these sources in the sample $n(D)$,
equation 
\eqref{eq:flucts-criterion-2}
can be rewritten in the following way, defining the quantity $\Upsilon$:
\begin{equation}
\Upsilon \equiv {\Delta\Omega\over \Omega} {dn\over df} \gg 1.
\label{eq:criterion-refined}
\end{equation}
The number of sources $n(D)$ is readily calculated from equation \eqref{eq:defNDM}:
\[
n(D) = \int_0^D dD D^2 \nu(D) L(M_*(D)). 
\]
The fractional flux $f(D) \equiv F(D) / F (D_{\rm max})$, where 
\[
F(D) ={F_0\over 4\pi}  \int_0^D dD \nu(D) L(M_*(D)) w(D) J(D),
\]
represents the total flux from sources closer than $D$. Note the
appearance of the weights $w(D)$ assigned by the sliding-box method in the
last equation.
Neglecting deviations of $\nu(D)$ from 1,
which is reasonable on cosmological scales,
we use the above equations to find that
\begin{equation}
\label{eq:dndf_slidbox}
{dn\over df} \simeq {3 N_V(D)\over D J(D)}\int_0^{D_{\rm max}} dD J(D) \quad \textrm{(sliding box),}
\end{equation}
where $N_V(D) = N(D,M_*(D))$ is the number of galaxies in the
volume-limited sample at distance $D$ (cf. eqs.~(\ref{eq:defNDM})).
We stress that equation~\eqref{eq:dndf_slidbox} is valid for fluxes
modeled using the sliding-box technique. For a single volume-limited subsample
that is valid up to $D_{\rm max}$, a similar computation yields:
\begin{equation}
\label{eq:dndf_vlim}
{dn\over df} \simeq {3 N_V(D_{\rm max}) D^2 \over D_{\rm max}^3 J(D)}\int_0^{D_{\rm max}} dD J(D) 
\quad \textrm{(vol. ltd.),}
\end{equation}
Comparing eqs~\eqref{eq:dndf_slidbox} and \eqref{eq:dndf_vlim}, we see that
the number of sources contributing to a given flux fraction at distance $D$
is increased by a factor $N_V (D) D_{\rm max}^3 / N_V (D_{\rm max}) D^3$. This
factor is unity at $D=D_{\rm max}$ (where the sliding-box method offers no
improvement), but may become very large at small distances.

Inserting equation~\eqref{eq:dndf_slidbox} into 
\eqref{eq:criterion-refined} 
brings us to the final expression for our criterion of small fluctuations:
\begin{equation}
\Upsilon \simeq 
{3 \Delta\Omega \over \Omega}
{N_V(D) D_{\rm max}\over D}  {\int_0^{D_{\rm max}} dD J(D)\over D_{\rm max} J(D)}
 \gg 1\, .
\label{eq:criterion-final}
\end{equation}
The three factors in equation~\eqref{eq:criterion-final} respectively encode
the dependence of $\Upsilon$ on the angular scale, on the statistics
of the flux-limited sample,
and on the attenuation of the model flux. The last factor reduces to
unity in the case of no attenuation. 
The equation has to hold for all values of $D$; when it is violated
$\mathcal{O}(1)$ fluctuations in flux may occur in regions of angular
size $\Delta\Omega$ due to the contribution of a single source.

We now discuss some of the quantities entering equation~\eqref{eq:criterion-final}.
The number of galaxies $N_V(D)$ becomes small at both very large and very small
distances, potentially leading to large fluctuations.
For large distances this can be prevented by considering
only sources up to a maximum distance $D_{\rm max}$ and 
assuming an isotropic flux from sources beyond that distance. Alternatively,
particle horizons may provide a natural maximum distance (see below).
In the case of small distances, 
the number of nearby sources is small while their contribution may be
important due to their proximity. Unlike the fluctuations at large
distances which are due to our poor knowledge of the galaxy
distribution at those scales, the fluctuations at small distances are
physical and may represent the actual flux variations due to close
sources. Their complete treatment may require a case-by-case study of
the most nearby objects.

\begin{figure}
\includegraphics[angle=270, width=8cm]{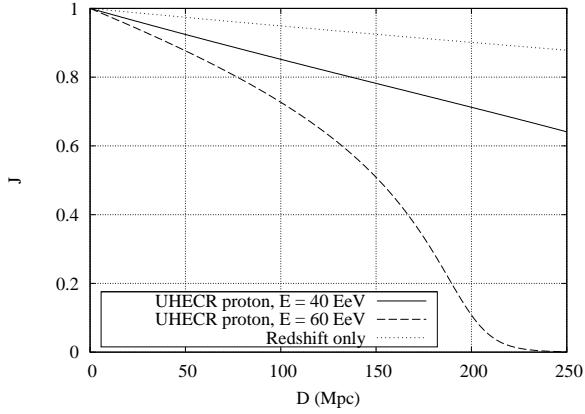}
\caption{\label{fig:ffunc} Flux suppression factor $J$ as a function
of distance for three different scenarios: redshift only,
UHECR protons with energy above 40 EeV, and 
UHECR protons with energy above 60 EeV. In producing this figure we have assumed
a power-law injection spectrum with index $p=2.2$ extending to very high energies.}
\end{figure}

\begin{figure}
\includegraphics[angle=270, width=8cm]{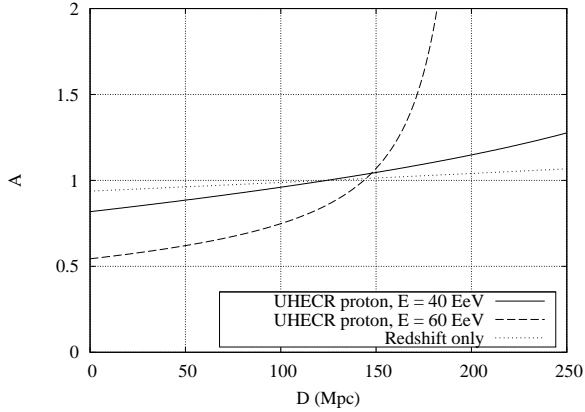}
\caption{\label{fig:attfactor}
Flux attenuation factor $A$ defined in eq.~\eqref{eq:defA} as a function
of distance for 
the same scenarios as shown in
figure~\ref{fig:ffunc}.}
\end{figure}

The last factor in equation~\eqref{eq:criterion-final} encodes the effect
of flux attenuation, which can play an important role in modeling the flux of 
UHECRs and of very-high energy gamma rays. Focusing on
the case of UHECRs, we show in figure \ref{fig:ffunc} the flux
attenuation factor $J$ as a function of distance for UHECR protons.
The attenuation factor is obtained using a numerical cosmic-ray propagation
code described in \citet{Koers:2008hv,Koers:2008ba}.
For comparison the attenuation factor due to redshift only is also shown
in the figure. 
In figure \ref{fig:attfactor} we show the quantity
\begin{equation}
\label{eq:defA}
A \equiv {\int_0^{D_{\rm max}} dD J(D)\over D_{\rm max} J(D)} \, ,
\end{equation}
which accounts for flux attenuation in equation~\eqref{eq:criterion-final}. 
Note that,
as indicated in figure \ref{fig:ffunc},
the horizon for UHECR protons
above 60 EeV is around 200 Mpc. Since sources beyond this distance do not
contribute to the observed flux, the requirement $\Upsilon \gg 1$
should be satisfied automatically. We observe from
figure \ref{fig:attfactor}
that $A$ indeed blows up around 200 Mpc, which guarantees that $\Upsilon \gg 1$
for any value of $\Delta \Omega$ or $N_V$.

\section{Example: UHECR flux predictions using the 2MRS catalog}
\label{section:2MRS}
In this section we apply the sliding-box technique to model the flux of
UHECR protons with energies in excess of 60 EeV from
sources tracing the distribution of matter in the Universe.
In modeling the effect of flux suppression due to attenuation, we 
assume a power-law injection spectrum with index $p=2.2$
extending to very high energies.
For comparison we also model the flux distribution using
a single volume-limited
galaxy sample up to 250 Mpc.

\begin{figure}
\includegraphics[angle=270, width=8cm]{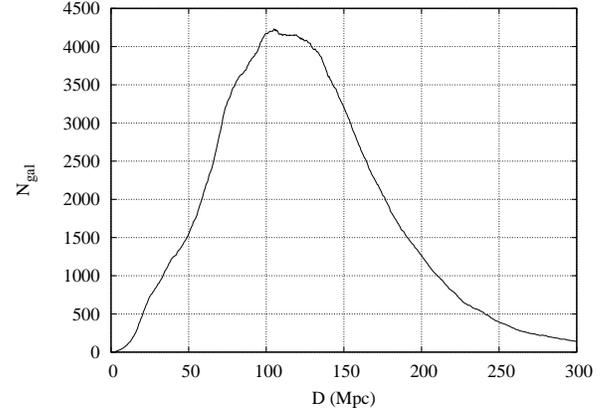}
\caption{\label{fig:Nvlim} Size of a volume-limited sample of our 2MRS sample
up to distance $D$.}
\end{figure}

The distribution of matter is modeled using a subset of the 2MRS galaxy sample.
This subset does not cover the galactic plane with
$|b|<10\degree$, nor the region with $\delta < -30\degree$
(see footnote \ref{footnote1}).
Due to these cuts the catalog covers $63\%$ of the sky, so that
the total field of view is $\Omega = 0.63  \, \cdot \,  4 \pi = 7.9$ srad.
In figure~\ref{fig:Nvlim} we show
the number of galaxies in a volume-limited sample
$N_V (D)$ of the 2MRS as a function of distance $D$.

In actual flux computations, dividing the sky into bins of fixed size
has disadvantages related to boundary effects and 
the arbitrariness of the binning scheme. These problems are avoided with 
an angular smearing routine, which essentially
replaces the point-source flux of an individual source
by a (Gaussian) probability distribution. Adopting a smearing routine,
the flux in a given
direction $\vec{n}$ is computed as follows:
\begin{equation}
\label{eq:totalfluxinbin}
\Phi (\vec{n}) = \sum_i \phi_i (\theta) \, ,
\end{equation}
where
\begin{equation}
\label{eq:Gaussiansmearing}
\phi_i (\theta) = \frac{F_i \exp( -\theta^2/\theta_{\rm s}^2)}{\pi \theta_s^2} \, .
\end{equation}
Here $F_i$ represents the flux from galaxy $i$,
$\theta$ denotes the angle between the galaxy and the line of sight
$\vec{n}$, and $\theta_{\rm s}$ is the smearing angle.

\begin{figure}
\includegraphics[width=8cm]{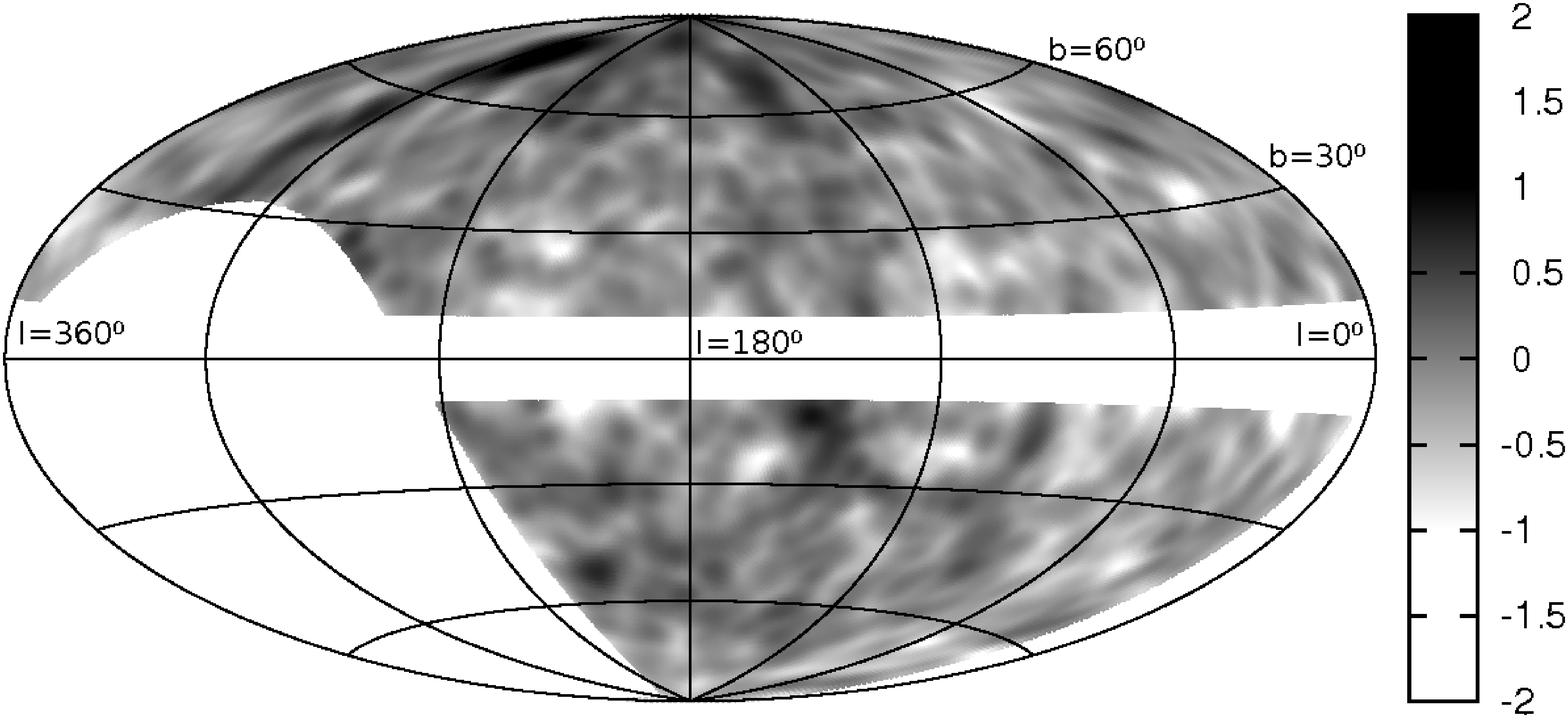}
\vspace{0.2cm}

\includegraphics[width=8cm]{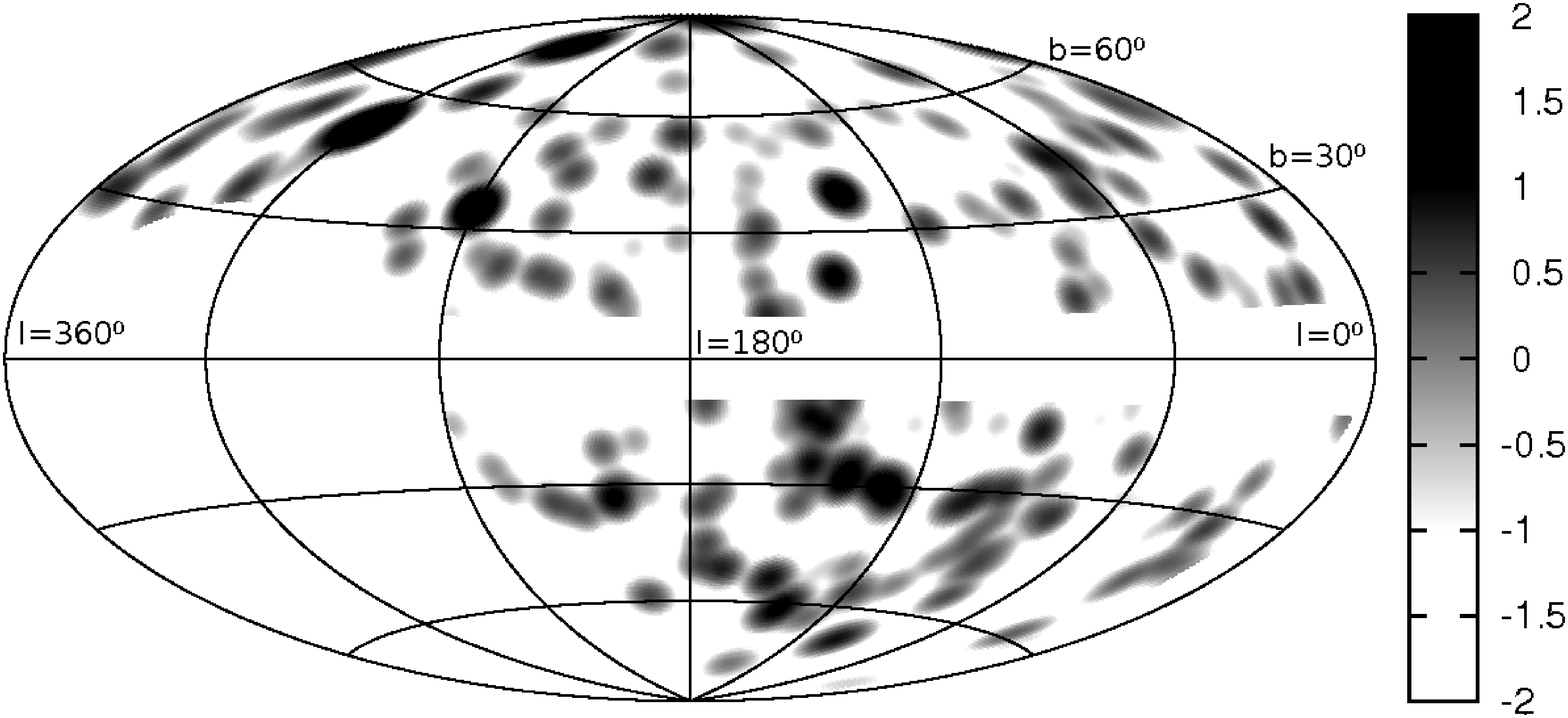}
\caption{\label{fig:skymaps} Aitoff projection of the sky in galactic
coordinates showing the model flux of UHECR protons above 60 EeV from sources
tracing the distribution of galaxies up to 250 Mpc. The grayscale
shows the relative flux on a logarithmic scale. (Areas in white are
not covered by our subsample of the 2MRS catalog.) The top panel shows
the flux constructed from the original flux-limited sample using the
sliding-box technique; the bottom panel shows the flux constructed
from a volume-limited subsample up to 250 Mpc.  For both cases we have
removed sources closer than 5 Mpc and smeared the flux distribution
with $\theta_{\rm s} = 3\degree$.}
\end{figure}

In figure \ref{fig:skymaps} we show  model UHECR flux maps
obtained with equation~\eqref{eq:totalfluxinbin} using the full 2MRS sample
with the sliding-box
method (top panel) and, for comparison, using  
a volume-limited subsample  extending to 250 Mpc (bottom panel).
A comparison between the two panels demonstrates the  significant 
increase in accuracy achieved
with the sliding-box technique.  In particular, the shot noise
artefacts that are visible in the bottom panel are absent
in the top panel.

\begin{figure}
\includegraphics[angle=270, width=8cm]{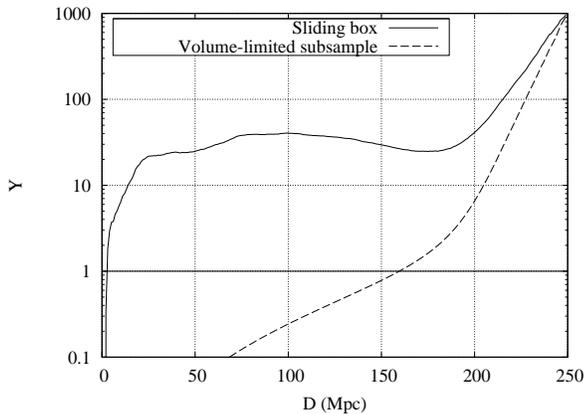}
\caption{\label{fig:limits} $\Upsilon$ as a function of $D$
for the case of 60 EeV UHECR protons 
modeled using our subset of the 2MRS catalog.
The figure applies to a smearing angle $\theta_{\rm s} = 3\degree$,
corresponding to $\Delta \Omega = 0.01$  srad;
the total field of view $\Omega = 7.9$ srad.}
\end{figure}

We now consider the fluctuations associated with the
finite number of galaxies in the 2MRS catalog. 
In figure~\ref{fig:limits}  we show the 
quantity $\Upsilon$ defined in
eq.~\eqref{eq:criterion-final} as a function of $D$
for the exemplary case of 60 EeV UHECR protons. For comparison
we also show $\Upsilon$ for the volume-limited sample
(using eqs.~\eqref{eq:criterion-refined}
and \eqref{eq:dndf_vlim}).
Because 63\%
of the total flux is contained within the opening angle
$\theta_{\rm s}$, the solid
angle $\Delta \Omega$ that enters in eq.~\eqref{eq:criterion-final}
is related to $\theta_{\rm s}$ as follows:
\begin{equation}
\label{DOmsmear}
\Delta \Omega  =  \frac{2 \pi (1-\cos \theta_{\rm s})}{0.63}
\simeq 5 \, \theta_{\rm s}^2  \,  .
\end{equation}
We observe from figure~\ref{fig:limits}
that, for the sliding-box method, $\Upsilon \gg 1$ for distance between 5 and 250
Mpc. Hence the results of the previous section imply
that fluctuations associated with the finiteness of the galaxy sample
should be small. For the volume limited subsample, on the other hand,
the figure indicates that strong fluctuations are to be expected.

\begin{figure}
\includegraphics[angle=270, width=8cm]{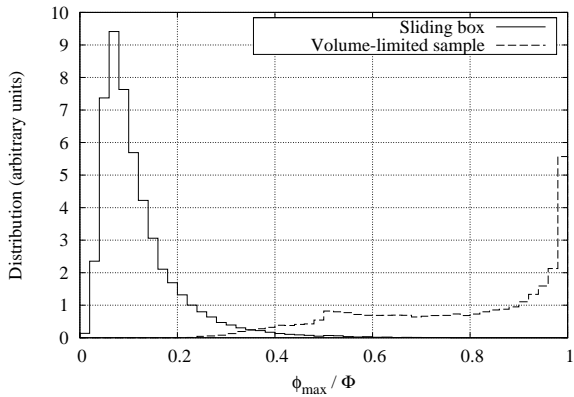}
\caption{\label{fig:distfi} Distribution of $ F_{\rm max} / F_{\Delta \Omega}$, i.e.
the ratio of the largest individual contribution to the total flux within
solid angle $\Delta \Omega$. This figure applies to the flux map
shown in the top panel of figure~\ref{fig:skymaps}.
}
\end{figure}

The estimates on the strength of fluctuations can be
verified through a direct computation of the ratio of
individual flux contributions $\phi_i$ to the total flux $\Phi$.
Sampling over many directions on the sky, we compute the total flux  $\Phi$
via equation \eqref{eq:totalfluxinbin} and keep track of 
$\phi_{\rm max} = \max \phi_i$, the largest
individual contribution to $\Phi$.
The distribution of $\phi_{\rm max} / \Phi$
is shown in figure~\ref{fig:distfi}
for the flux
maps shown in figure~\ref{fig:skymaps}, i.e.
for the case of
UHECR protons with energy in excess of 60 EeV and smearing angle
$\theta_{\rm s}=3\degree$.
As may be verified from figure~\ref{fig:distfi},
$\phi_{\rm max} / \Phi \ll 1$ for the sliding-box method: no single source
outshines the bulk.
On the other hand, if we model the flux distribution from
the volume-limited sample (bottom panel of figure~\ref{fig:skymaps}),
we find that the distribution of
$\phi_{\rm max} / \Phi$ peaks near 1. In this case
$\mathcal{O} (1)$ fluctuations in the predicted flux
due to a single source are common (which is also clear from
the bottom panel of figure~\ref{fig:skymaps}), which means that
the galaxy sample is too small to provide an accurate flux map.

\section{Summary}
\label{sec:summary}

We have addressed the problem of flux calculation from sources
tracing the galaxy distribution in the Universe.  We have discussed a
sliding-box weighting scheme, building on the work of
\citet{1971MNRAS.155...95L}, that makes use of the
information contained in a flux-limited galaxy catalog
in the most efficient way.  This
scheme allows us to represent the distribution of matter up to large
scales while representing the distribution of matter on small scales
with maximum accuracy.  The resulting weight function is related to
the galaxy luminosity function and may be used to infer the latter
under the assumption of its coordinate independence.

The sliding-box weighting scheme suppresses efficiently fluctuations
due to the finite size of the sample at most distances except the
largest and the smallest ones. We have presented estimates on the size
of the remaining fluctuations. These estimates can be used to
determine a maximum distance at which the catalog should be cut, or to find
the minimum angular scale on which flux maps can be constructed accurately.
We would like to stress that our estimates regard the
\emph{size} of fluctuations, and not their \emph{importance}.  For
example, in a statistical test based on model flux distributions,
fluctuations of order unity may be acceptable if the overall flux
distribution shows very strong contrasts or when the angular scales of
interest are much larger than the scale at which fluctuations occur.

An advantage of the sliding-box scheme is that it allows a
straightforward generalization to the cases when the sources trace
preferentially certain types of galaxies, or when the source
luminosity is correlated with the galaxy type. Such effects may be
accounted for by pre-weighting the galaxies in the catalog in a
corresponding way and modifying accordingly the sliding-box weighting
scheme.

\section*{acknowledgments}
We gratefully acknowledge the use of the 2MRS catalog, which was provided
to us by John Huchra. We thank John Huchra and Tom Jarrett for useful discussions,
and the anonymous referee for constructive comments.
H.K. and P.T. are supported by Belgian Science Policy under IUAP VI/11
and by IISN. The work of P.T. is supported in part by the
FNRS, contract 1.5.335.08.

\bibliographystyle{mn2e}
\bibliography{refs}

\label{lastpage}

\end{document}